\newcommand\CPX{Cu(py)$_2$(Cl$_{1-x}$Br$_{x}$)$_2$}

\newcommand\CPY{Cu(py)$_2$(Br$_{1-y}$Cl$_{y}$)$_2$}
\newcommand\CPC{Cu(py)$_2$Cl$_{2}$}
\newcommand\CPB{Cu(py)$_2$Br$_{2}$}
\newcommand\msr{$\mu$SR}

\documentclass[prl,twocolumn,floatfix,preprintnumbers,amsmath,amssymb,superscriptaddress]{revtex4}

\usepackage{graphicx}
\usepackage{color}

\usepackage{dcolumn}
\usepackage{bm}

\begin{document}

\title{Ordering in weakly coupled random singlet spin chains}

\author{M. Thede}
\affiliation{Neutron Scattering and Magnetism, Laboratory for Solid State
Physics, ETH Z\"urich, Z\"urich, Switzerland} \affiliation{Laboratory for Muon Spin Spectroscopy, Paul Scherrer Insitut,
Villigen-PSI, Switzerland}
\author{F. Xiao}
\affiliation{Neutron Scattering and Magnetism, Laboratory for Solid State
Physics, ETH Z\"urich, Z\"urich, Switzerland}
\affiliation{Department of Physics, Clark University, 950 Main St.,
Worcester, MA 01610, USA.}
\author{Ch. Baines}
\affiliation{Laboratory for Muon Spin Spectroscopy, Paul Scherrer Insitut,
Villigen-PSI, Switzerland}
\author{C. Landee}
\affiliation{Department of Physics, Clark University, 950 Main St.,
Worcester, MA 01610, USA.}
\author{E. Morenzoni}
\affiliation{Laboratory for Muon Spin Spectroscopy, Paul Scherrer Insitut,
Villigen-PSI, Switzerland}
\author{A. Zheludev}
 \email{zhelud@ethz.ch}
 \homepage{http://http://www.neutron.ethz.ch/}
\affiliation{Neutron Scattering and Magnetism, Laboratory for Solid State
Physics, ETH Z\"urich, Z\"urich, Switzerland}

\date{\today}

\begin{abstract}
The influence of bond randomness on long range magnetic ordering in the weakly
coupled $S=1/2$ antiferromagnetic spin chain materials \CPX\ is studied by
muon spin rotation and bulk measurements. Disorder is found to have a strong effect on
the ordering temperature $T_\mathrm{N}$, and an even stronger one on the saturation magnetization $m_0$,
but considerably more so in the effectively lower-dimensional
Br-rich materials. The observed behavior is attributed to Random Singlet ground states
of individual spin chains, but remains in contradiction with
chain mean field theory \cite{Joshi2003} predictions. In
this context, we discuss the possibility of a universal {\it
distribution} of ordered moments in the weakly coupled Random Singlet chains
model.
\end{abstract}

 \maketitle

The ground states of unfrustrated classical systems are typically robust with
respect to weak Hamiltonian disorder. A case in point is the classical
Heisenberg antiferromagnet (HAF). Randomizing the strength (but not the
signs) of exchange interactions leaves the fully aligned Neel ground state
completely intact. In contrast, in quantum systems, arbitrarily weak disorder
will modulate the strengths of local quantum fluctuation and often
qualitatively reconstruct the ground state. The one-dimensional quantum
$S=1/2$ HAF is an extreme example. For uniform chains, the ground state is a
Tomonaga-Luttinger spin liquid (TLSL) \cite{Giamarchibook}. The introduction
of arbitrary weak bond randomness gives rise to  the so-called Random Singlet
(RS) phase \cite{Ma1979,Dasgupta1980,Doty1992,Fisher1994}. In the RS state,
spin correlations are protected from localization effects by particle-hole symmetry \cite{Theodorou1976,Giamarchibook}.
Nevertheless, the scaling laws \cite{Fisher1994,Damle2000,Motrunich2001}, although universal and
independent of the details of disorder,
are markedly different from those of the TLSL. In experiments on real materials,
one has to deal with quasi-one-dimensional (quasi-1D) spin systems. A
divergent correlation length in individual chains ensures 3D long range order
at $T_\mathrm{N}>0$ for arbitrary weak inter-chain interactions $J'$, in both
the disorder-free \cite{Scalapino1975} and disordered cases
\cite{Fisher1994,Joshi2003}. An intriguing question is to what extent
the peculiarities of the RS state in $d=1$ translate into unusual features of
the ordered phase in $d=3$.

To date, RS-forming bond randomness in quasi-1D magnets has received
considerably less attention than dilution-type disorder due to spin
substitution \cite{Birgeneau1980,Endoh1981,Uchiyama1999}. For the former, the
existing predictions are derived from chain-Mean Field (chain-MF) theory
\cite{Joshi2003}. The main result is that random bonds tend to {\it increase}
both $T_\mathrm{N}$ and the ordered moment $m_0$ at $T\rightarrow 0$. This peculiar ``order from disorder''
effect is related to an abundance of very loosely coupled and almost free
spins in the RS state of isolated chains \cite{Ma1979,Doty1992}. In origin, it is similar to
disorder-induced ordering in frustrated magnets, where spin fluctuations are also strong \cite{Henley1989,Chubukov1992}. In coupled
random chains, for weak $J'$, one gets $T_\mathrm{N}\propto J'm_0$ \cite{Joshi2003}. In contrast, for
disorder-free case, the in-chain interactions enter the relation
explicitly: $T_\mathrm{N}\propto J m_0^2$ \cite{Schulz1996}. On the
experimental side, the challenge is to measure the very small sublattice
magnetization that arises in the weak-coupling regime, where this theory may
be expected to apply. In the present work we overcome this difficulty by employing
the sensitive muon spin rotation (\msr) technique, which has emerged as a tool
of choice for the study of quantum magnetism \cite{Pratt2011}. We  directly measure the
relative variations of $m_0$ and $T_\mathrm{N}$ in the prototypical
bond-disordered quasi-1D $S=1/2$ HAF systems \CPX. The observed behavior,
while starkly different from that in disorder-free chains,
in apparent contradiction with chain-MF predictions for coupled RS chains.

\begin{figure}
\includegraphics[width=\columnwidth]{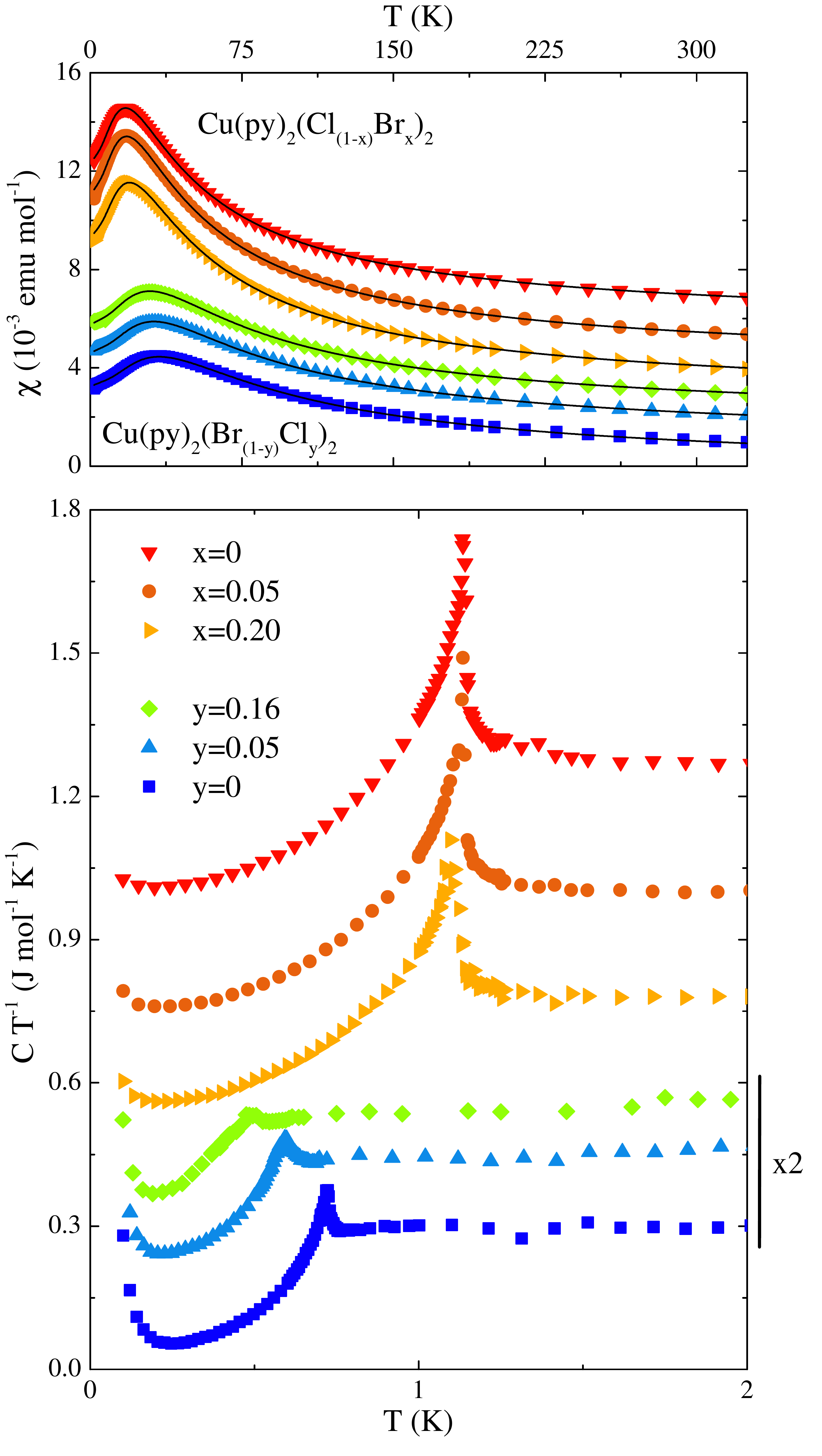}
\caption{(Color online) Bulk properties of \CPX\  samples. (a) Measured temperature dependence
of magnetic susceptibility for a field applied along the chain axis (symbols),
and fit of the theoretical curve for the uniform quantum $S=1/2$ HAF chain
\cite{Bonner1964,Johnston2000} (solid line). From the buttom up, the data are offset
along the $y$ axis by 0, 1, 2, 3, 4.2, and 5.8 10$^{-3}$~emu/mol, respectively.
(b) Measured temperature dependence of specific heat
(symbols). The vertical offsets are 0, 0.16, 0.27, 0.55, 0.75 and 1 J/mol K, respectively. \label{bulk}}
\end{figure}

Our target compounds are derivatives of \CPC,\ one of the first known and
extensively studied $S=1/2$ spin chain materials \cite{Endoh1974}. Single
crystal samples with varying Br content $x$ are straightforward to grow from
solution by slow evaporation. In \CPC\ (space group $P2_1/n$,
$a=16.967$\AA,\ $b=8.5596$\AA, $c=3.8479 $\AA,\  $\beta=91.98^\circ$) the
chains are formed by magnetic $S=1/2$ Cu$^{2+}$ ions linked by superexchange
bonds via the halogen sites. The temperature dependence of magnetic
susceptibility (Fig.~\ref{bulk}a) shows a broad Bonner-Fischer (BF) maximum
\cite{Bonner1964,Johnston2000}, characteristic of a quantum $S=1/2$ chain
with an AF exchange constant $J_{x=0}=2.35$~meV. The chain-structure of \CPB\
is quite similar ( $P2_1/n$, $a=8.424$\AA,\ $b=17.599$\AA, $c=4.0504$\AA,\
$\beta=97.12 ^\circ$ \cite{Morosin1975}), as are the measured magnetic
susceptibility curves. However,  the
in-chain exchange constant is larger: $J_{x=1}=4.58$~meV. Weak inter-chain
interactions lead to 3D ordering in both materials, at $T_\mathrm{N}=1.15$~K
\cite{Takeda1973} and $T_\mathrm{N}=0.72$~K for \CPC\ and \CPB,\
respectively. The transitions are marked by well-defined lambda-anomalies in
the measured temperature dependence of specific heat $C(T)$, as shown in
Fig.~\ref{bulk}b \footnote{All specific heat data shown in this work were
measured on single crystals using a Quantum Design Physical Properties
Measurement system PPMS-XL with a $^3$He-$^4$He dilution refrigerator
insert.}. Knowing $T_\mathrm{N}$ and $J$, allows us to estimate the effective
inter-chain coupling constants \cite{Yasuda2005}: $J'_{x=0}=0.05$~meV and
$J'_{x=1}=0.03$~meV for the two materials, correspondingly. The bromide is
clearly a much more one-dimensional system. It is useful to estimate the ordered moment at
$T\rightarrow 0$. Based on chain-MF results, for \CPC\ and \CPB\ we get
$m_{0,x=0}=0.15~\mu_\mathrm{B}$ and $m_{0,x=1}=0.08~\mu_\mathrm{B}$,
respectively. The smaller ordered moment, and hence a greater one-dimensional character of the
Br system is also manifested in the much weaker $C(T)$ lambda anomaly.

As determined by single crystal X-ray diffraction and chemical elemental
analysis, the  structures of \CPC\ and \CPB\ are stable with respect to
chemical substitution of Cl for Br and vice versa, for $x<0.4$ and $x>0.6$,
respectively. Id addition to changing bond angles due to a difference in ionic radii,
the more extended wavefunctions in Br$^{-}$ provides a stronger superexchange pathway compared to Cl$^{-}$, typically by a factor of 2 to 4 \cite{Willet1986}.
This strategy of creating bond-disordered systems has been
previous successfully applied in other materials such as
IPA-Cu(Cl$_{1-x}$Br$_x$)$_3$ \cite{Hong2010PRBRC},
piperazinium-Cu$_2$(Cl$_{1-x}$Br$_x$)$_6$ \cite{Huevonen2012},
H$_8$C$_4$SO$_2\cdot$Cu$_2$(Cl$_{1-x}$Br$_x$)$_4$ \cite{Wulf2011}  and
NiCl$_2\cdot$4SC(NH$_2$)$_2$ \cite{Yu2011}. Due to the slightly different
structures of the parent compounds, we are actually dealing with {\it two}
tunable random-bond spin chain materials, on the Cl-rich and Br-rich ends of
the \CPX\ line, respectively. We will reserve the formula
\CPX\ for Cl-rich compounds ($x<0.5$) and use \CPY\ to denote materials on
the Br end ($y<0.5$).

\begin{figure}
\includegraphics[width=\columnwidth]{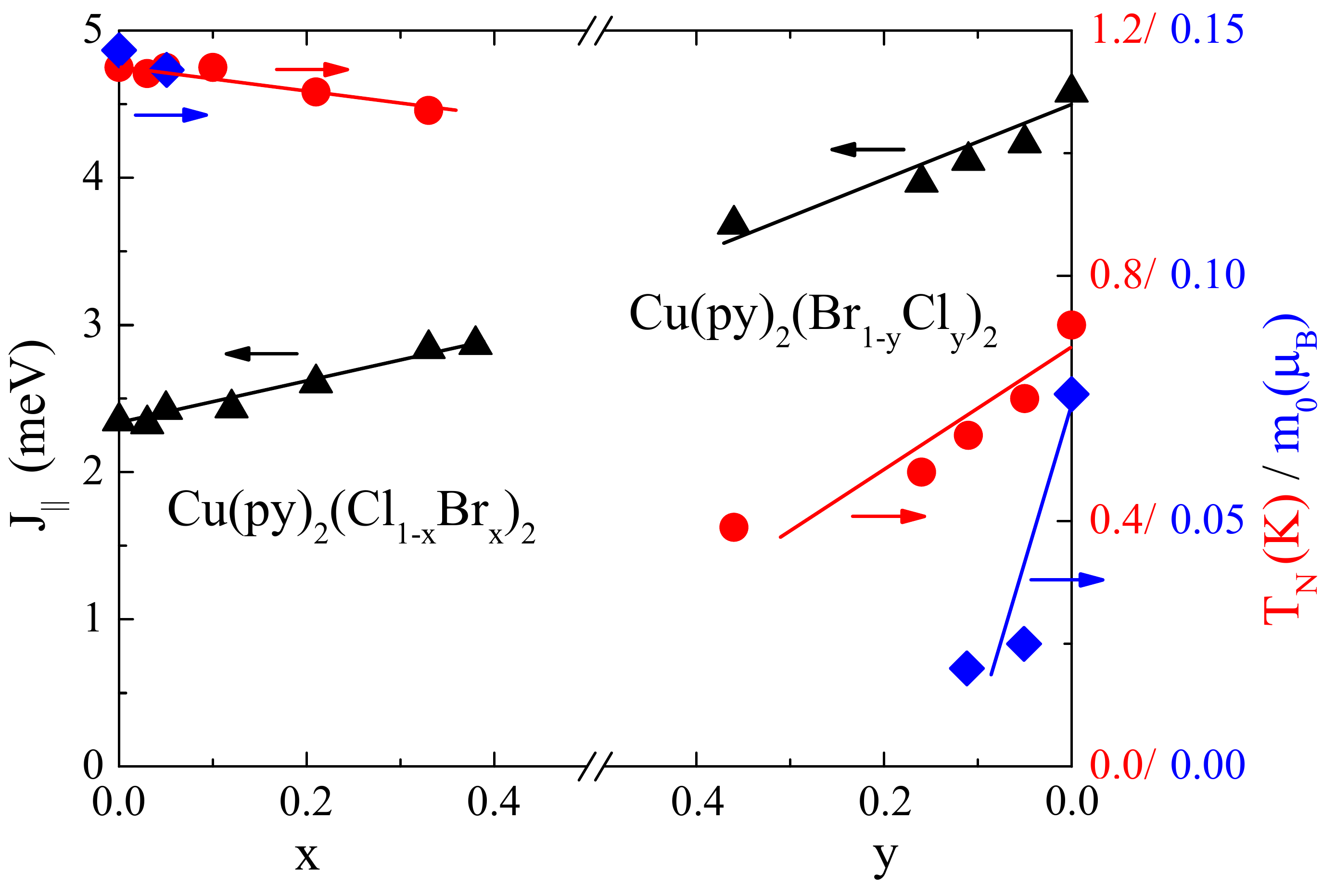}
\caption{(Color online) Composition dependence of the in-chain exchange constant $J$ (from
fits to $\chi(T)$ ), the ordering temperature $T_\mathrm{N}$ (from calorimetry)
and the low-temperature ordered moment $m_0$ (chain-MF estimates and $\mu$-SR measurements). \label{vsxy}}
\end{figure}

The bulk properties of the halogen-disordered samples resemble those of the
corresponding disorder-free systems. Typical measured magnetic susceptibility
data are plotted symbols in Fig.~\ref{bulk}. The derived average in-chain exchange constant $J$ steadily increases  with increasing
Br content. A 3D magnetic ordering transition is observed at low temperatures
in all composition  studied(Fig.~\ref{bulk}, lower panel). $T_\mathrm{N}$ decreases with
Br content on the Cl-rich end and with Cl concentration in Br-rich samples
(Fig.~\ref{vsxy}). The variation is more pronounced in \CPY. Moreover, the
corresponding lambda-anomaly weakens and slightly broadens with increasing Cl
content in \CPY, while it remains almost unchanged in \CPX. The measured transition temperature and average exchange
constant are plotted versus composition in Fig.~\ref{vsxy}.

\begin{figure}
\includegraphics[width=\columnwidth]{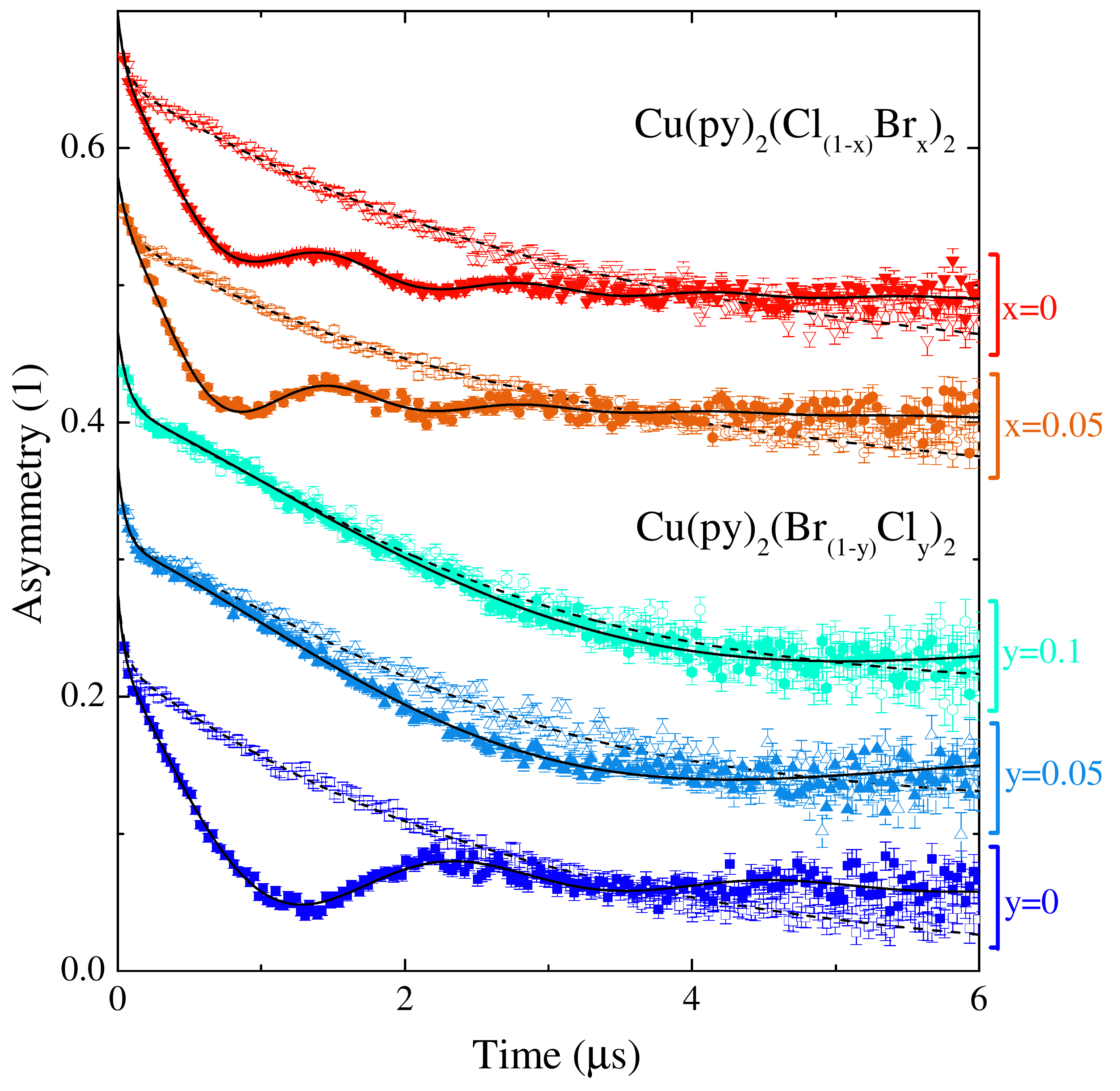}
\caption{(Color online) Muon spin asymmetry measured vs.\ time in several \CPX\ and \CPY\
samples at $T\sim 20$~mK (solid symbols) and just above $T_\mathrm{N}$ (open
symbols). The plots for $y=0$, $0.05$ and $0.1$ and $x=0.05$, $0$ are offset
along the $y$ axis by 0, 0.1, 0.2, 0.3 and 0.43, respectively. The solid lines are fits to
the data as described in the text. \label{spectra}}
\end{figure}

The main focus of the present study is on $\mu$-SR measurements. This
technique has been instrumental in the study of very small static\cite{Lancaster2006,Sugano2010} and dynamic\cite{Chakalin2003,Pratt2006} moments
in spin chain systems.  It probes
the local magnetic fields at the stopping sites of muons implanted into the
sample \cite{YaouancReotier}. These, in turn, are expected to be proportional
to the static ordered moment. We performed measurements on powder samples of
\CPX\ and \CPY\ with $x=0,0.05$ and $y=0$, $0.05$ and $0.1$ at the LTF spectrometer at the S$\mu$S muon source at Paul Scherrer Institut.
Typical muon spin relaxation/rotation curves measured in zero applied field (ZF) are shown in Fig.~\ref{spectra}.
Several distinct time scales are apparent. In all samples, a very rapid decay
at short times can be attributed to muonium formation with the organic ligand
\cite{Rhodes2001}. This contribution appears to be temperature-independent and is not directly relevant to the physics discussed here. At the largest
time scales, a slowly decaying tail is due to the non-precessing muon spin components
parallel to the local field, and to muons stopping outside the
sample \cite{YaouancReotier}.

As expected, in the paramagnetic phases of all samples one only observes a
slow decay of muon polarization, with no oscillatory behavior. Representative
data collected at $T>T_\mathrm{N}$ are shown in open symbols in
Fig.~\ref{spectra}. They can be modeled with exponential decay processes, as shown in dashed lines. We conclude that above the ordering
transition, all materials studied behave very similarly, despite the
different levels of disorder.

\begin{figure}
\includegraphics[width=\columnwidth]{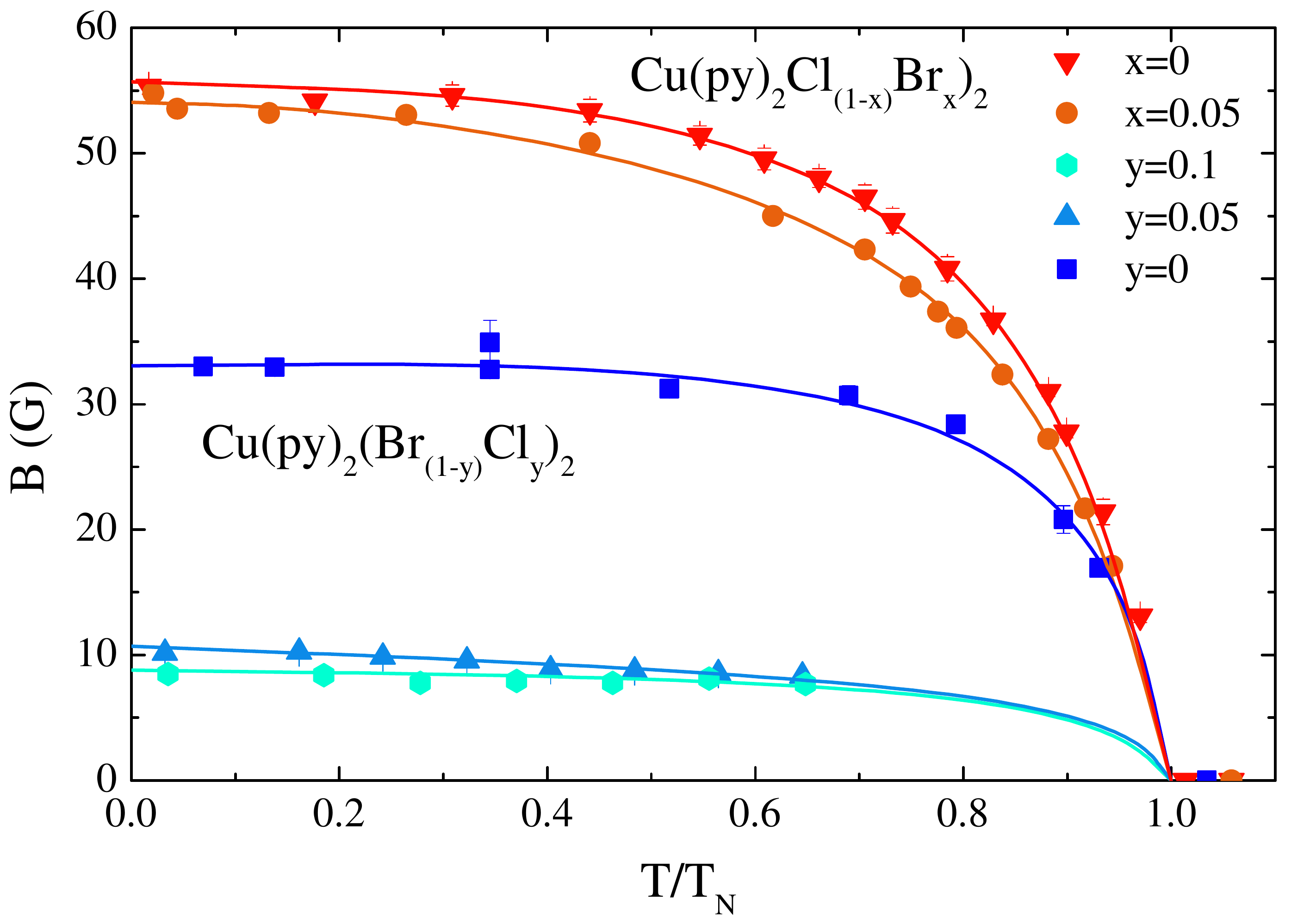}
\caption{(Color online) Characteristic field at the principal muon
stopping site, assumed toi be proportional to the magnetic order parameter,  plotted against temperature in units of the average in-chain
exchange constant $J$ in \CPX\ and \CPY\ (symbols). Within each family of
materials, the relative magnitudes of the precession field can be directly
compared. Lines are guides to the eye.
\label{orderp}}
\end{figure}

Clear differences emerge at low temperatures, in the magnetically ordered
phase (Fig.~\ref{spectra}, solid symbols). Although the relevant time scales
turned out to be strongly dependent on composition, it is possible to provide
a common description of the low-temperature $\mu$-SR spectra in all samples.
In addition to the fast and slow background contributions described
above, for $T<T_\mathrm{N}$, the main effect is the appearance of spontaneous muon spin precession.
An application of a small longitudinal field
parallel to the muon spin direction recovered all polarization apart from the muonium contribution. This
observation shows that the observed spin relaxation and rotation are due to {\it static} internal fields,
which we attribute to ordered Cu$^{2+}$ moments.
We chose to model these processes as a sum of a
damped Bessel function and an exponential term. The spectra are then
described as:
\begin{eqnarray}
A(t) & = & A_1J_0(\omega t+\phi)\exp(-\lambda_1 t)+ A_2\exp(-\lambda_2 t)+
\nonumber
\\
 & + &A_\mathrm{fast}\exp(-\lambda_\mathrm{fast}t)+A_\mathrm{tail}\exp(-\lambda_\mathrm{tail}t).\label{order}
\end{eqnarray}

The first term on the RHS represents the precession of muons stopped at the
most probable sites \cite{Chakalin2003}. The choice of a Bessel function, typically used to
describe incommensurate structures \cite{YaouancReotier} is in our case
purely empirical. Nevertheless, its use is justified by preliminary neutron
diffraction evidence \footnote{M. Thede, unpublished (2012).} that the
magnetic structure is actually helimagnetic in the $b$ direction,
perpendicular to the chains. Its exponential envelope reflects a narrow Lorentzian
distribution of local fields and also takes into account the depolarization by
nuclear spins. The characteristic muon spin precession frequency $\omega$ is directly proportional to the magnitude of the static magnetic
field $\omega= \gamma B$, with $\gamma=85.16$~krad~s$^{-1}$G$^{-1}$, is thus our primary measure of the static magnetic order.

The second term in Eq.~\ref{order} empirically describes the multitude of
other stopping sites sensitive to the static magnetic order. For each sample,
the site occupancies $A_1$ and $A_2$, as well as parameters for ``fast'' and
``tail'' contributions, were determined in global fits to the data
collected in the entire temperature range. The parameter $\phi$ determines
the functional shape of the oscillatory term. At each temperature, it was
globally applied to all samples, separately on the Br-rich and Cl-rich ends,
to allow a direct comparison of the precession frequencies $\omega$ within
each family of materials. For each sample, the parameters $\omega$,
$\lambda_1$ and $\lambda_2$ were refined at each temperature separately. The
model provides excellent fits to all data collected for the ordered state in
all samples. Typical fits are shown in heavy solid lines in
Fig.~\ref{spectra}. The values of all fit parameters for all samples and
their temperature dependencies are deposited as supplementary material.

The main result of our analysis is the temperature dependence of the local field  $B$, plotted in Fig.~\ref{orderp}. For each composition, the
temperature axis has been normalized to the value of $T_c$ derived  from
calorimetric measurements. Three conclusions can be immediately drawn. First,
in apparent contradiction with chain-MF theory, both $T_\mathrm{N}$ and  the saturation
magnetization $m_0$ are {\it reduced} in disordered samples, on both composition ends.
Second, the Br-rich materials are
considerably more affected by disorder that the Cl-rich
systems. Here, the oscillatory term is dramatically slowed already at 5\%
substitution, and is almost completely overrun by relaxation effects at 10\%
Cl. Experimental conditions and data quality are very similar in all 5 samples, so the disappearance of
clear oscillations is direct proof of $m_0$ reduction. This effect is fully consistent with
a progressive weakening of the $C(T)$ lambda anomaly in \CPY\ with increasing $y$ (Fig.~\ref{bulk}).
Third, for Br-rich samples the effect of bond disorder on  $m_0$ is much more drastic than on $T_\mathrm{N}$.
This is made particularly clear by the $m_0$ vs. composition
plot in Fig.~\ref{vsxy}. It is based on our initial estimates
for $m_0$ in the disorder-free materials, and on the assumption that within
each of the two series of materials $m_0$ is proportional to the Larmor field
$B$ extrapolated to zero temperature.

That disorder effects are not apparent in bulk
properties above $T_\mathrm{N}$ is not at all surprising. Recent numerical
simulations and experimental studies of the RS material BaCu$_2$SiGeO$_7$
\cite{Shiroka2011} have illustrated that the bulk effect of even very large
randomness may be very modest. Deviant behavior of susceptibility (in our case
measured down to 2~K) may emerge only at temperature
that are an order of magnitude smaller than the width of the bond probability
distribution $P(J)$. Even if we assume that in \CPY\ we are dealing with a clearly exaggerated bimodal distribution, a 10\%
substitution will produce a standard deviation of only about 0.7~meV. In this
conservative estimate, RS behavior may be expected to affect the bulk
properties only at temperatures below $\sim 1$~K. As far as the specific heat is concerned, for
RS chains one expects a power-law behavior with $C(T)\propto T^{\gamma_\mathrm{C}}$ \cite{Ma1979} The exponents
depends on the actual $P(J)$, and shows a slow
temperature dependence. For the relevant temperature range in our experiments ($T_\mathrm{N}<T\lesssim 5$~K),
for $P(J)$ with a support removed from $J=0$ (most certainly true in our case),
Ref.~\cite{Ma1979} suggests $0.9<\gamma_\mathrm{C}<1.1$. This explains why, despite the disorder, $C(T)/T$ remains roughly constant above $T_\mathrm{N}$ in all of our samples.

A key point is that for {\it 3-dimensional ordering} in at least for the Br-rich materials,  disorder  {\it has to be relevant}.
Specifically,  for $y=0.05$, from the measured values
of $J$ and $T_\mathrm{N}$, from chain-MF theory for {\it disorder-free} chains \cite{Schulz1996} one gets $m_0\approx0.06$~$\mu_\mathrm{B}$,
as compared to the much smaller the observed value $m_0\approx0.025$~$\mu_\mathrm{B}$. The discrepancy
can not be explained without invoking disorder effects, and
is even more drastic for $y=0.1$: $m_0\approx0.06$~$\mu_\mathrm{B}$ or a disorder-free model vs. $m_0<0.02$~$\mu_\mathrm{B}$ observed.
In contrast, the mismatch between $T_\mathrm{N}$ and $m_0$ in \CPX, as compared
to expectations for disorder-free chains, is not as drastic.
To explain this, we recall that RS properties emerge only below a certain
energy scale that is non-universal and depends on the initial
distribution of exchange constants \cite{Ma1979,Dasgupta1980,Doty1992,Fisher1994}.
In  \CPX, where the relative strength of inter-chain interaction is roughly 4 times stronger than in \CPY,
this energy scale may be much  lower than 3D interactions, making disorder irrelevant in the ordered phase.

The discrepancy between our findings and the chain-MF theory of Ref.~\cite{Joshi2003}  remains
to be explained. One tantalizing possibility is that the chain-MF approach may be in principle
be inapplicable to the RS phase \footnote{T. Giamarchi, private communication.}.
The latter features an abundance of weakly-dimerized spin degrees of freedom
for which quantum correlations with similar objects in adjacent chains simply
can not be ignored. A final note concerns the {\it homogeneity} of the static ordered moment. The
low energy physics of the RS phase is {\it exactly} that of non-interacting
random dimers with a universal probability distribution of dimer strength. These singlets will be partially polarized by the mean
exchange field. The degree of polarization will be determined by the strength
of the dimers. As a result, we expect a universal probability and spatial
{\it distribution} of static ordered moments $P(m)$ in the $T\rightarrow 0$
limit. Even though muon spectroscopy could, in principle, measure this
distribution directly, in \CPX\ and \CPY\ such an experiment appears
very challenging. We believe that
this elegant idea of a {\it universal distribution of ordered moments in
weakly coupled RS chains} deserves more theoretical and experimental
attention, using new materials for $\mu$-SR and using alternative techniques
such as NMR.

This work is partially supported by the Swiss National Fund under project
2-77060-11 and through Collaborative Project 8 of MANEP. We would like to thank Mark Turnbull for helpful discussions of the single crystal growth and diffraction analysis.


\end{document}